\documentclass[preprint,aps,showpacs]{revtex4}
\usepackage[dvips]{graphics}
\usepackage{amssymb}
\usepackage{epsfig}
\usepackage{amsmath}
\usepackage{makeidx}

\makeindex

\begin{document}

\title{Comment to article "A light--hole exciton in a quantum dot" by Y.H.Huo et al, Nature Physics {\bf 10}, 46 (2014).}
\author{E.Tsitsishvili}
\address{Institute for Cybernetics, Tbilisi Technical University,  S. Euli 5,
0186,  Tbilisi Georgian Republic}

\draft

\begin{abstract}

The exciton ground state in strained quantum dots similar to those
fabricated in article specified in the title is shortly discussed
within a relevant model Hamiltonian. Some characteristics of the
light--hole exciton ground state reached in a dot under the
tensile biaxial strain appear to be sensitive to the strain
anisotropy breaking a purity of this state. It refers in
particular to a degree of the in--plane polarization of the
emission and the fine structure of the ground state.

\end{abstract}

\pacs{78.67.Hc, 73.21.La}

\maketitle

\draft

In recent paper "A light--hole exciton in a quantum dot" by Y.H.
Huo et al.\cite{Huo} is reported about a creation of the
light--hole exciton ground state by applying biaxial tensile
strain to an initially unstrained quantum dot. This conclusion is
based, in particular, on the observation of the $z$--polarized
line in emission spectra of strained dots - an obvious sign that
the ground state is "mainly light--hole" exciton. A degree of the
in--plane polarization anisotropy of the emission contains a
definite information about a "purity" of this state. The last item
is discussed by authors in Supplementary Section II.6 on a basis
of theoretical results from Ref.\cite{Tonin}. While we acknowledge
the calculation way of the linear polarization degree used in
Ref.\cite{Tonin}, an application of the particular results
received there (formulae (1) and (7)) to the scenario taking place
in Ref.\cite{Huo} appears to be questionable. In this comment we
simply would like to give some additions and specifications.

It is a question about the states of the heavy--hole (HH) and
light--hole (LH) exciton with total angular momentum projection
$|J_z|=1$. For the in--plane symmetrical dots, this (two--fold
degenerate) bright state is circularly polarized and therefore
contributes to unpolarized emission. Any perturbation providing a
coupling between the states with $J_z$ of different signs leads
generally to a formation of elliptically polarized states.
Considering the HH--LH coupling of this type, the low--energy
exciton states  have a form (for simplicity, we take the same
envelope function for the heavy and light hole)
\begin{eqnarray}\label{Ellipstate}
 \Psi_{L}^{(\pm)} \approx  \Bigl[w_{hh}\;
\Bigl|\mp\frac{1}{2};\pm\frac{3}{2}\rangle + w_{lh}
\;\Bigl|\mp\frac{1}{2};\mp\frac{1}{2}\rangle \Bigr]\;.
\end{eqnarray}
The (square) amplitude $|w_{hh(lh)}|^2 = 0.5 \Bigl[1 \pm
\Delta_{lh}^{(0)} \Bigl( \sqrt{(\Delta_{lh}^{(0)})^2 + 4
\rho^2}\Bigr)^{-1}\Bigr]$ (with the mixing amplitude $\rho$ and
the LH--HH splitting $\Delta_{lh}^{(0)}>0$) determines the exciton
character $ P_{hh(lh)} = |w_{hh(lh)}|^2$, that is the probability
for the exciton to be HH(LH) exciton. Evidently, the above ground
state is mainly of the heavy--hole type in a weak coupling limit,
where $\rho \ll \Delta_{lh}^{(0)}$, while both the HH and LH
exciton characters are of the same order at $\rho \gg
\Delta_{lh}^{(0)}$. Having in mind a contribution of  the bright
exciton to a recombination, a degree of the linear polarization of
the emission is expressed generally as
\begin{eqnarray}\label{degreeofpol}
C = \frac{2\; \sqrt{3 P_{hh} \;P_{lh}}}{3 P_{hh} + P_{lh}} \;.
\end{eqnarray}
Here it is considered that the recombination probability is three
times larger for the HH  than LH exciton, see e.g.\cite{Efros}.
From Eq.(\ref{degreeofpol}) follows that the linear polarization
is equal to zero in absence of the HH--LH mixing, as expected,
whereas it becomes close to unity in a strong coupling limit where
$\rho \gg \Delta_{lh}^{(0)}$. The above scenario is familiar for
conventional quantum dots showing the anisotropic effects of an
intrinsic nature and keeping the exciton ground state of the
heavy--hole type, see e.g.\cite{Leger} and Ref.\cite{Tonin} as
well.

\begin{figure}
\includegraphics[scale=.3]{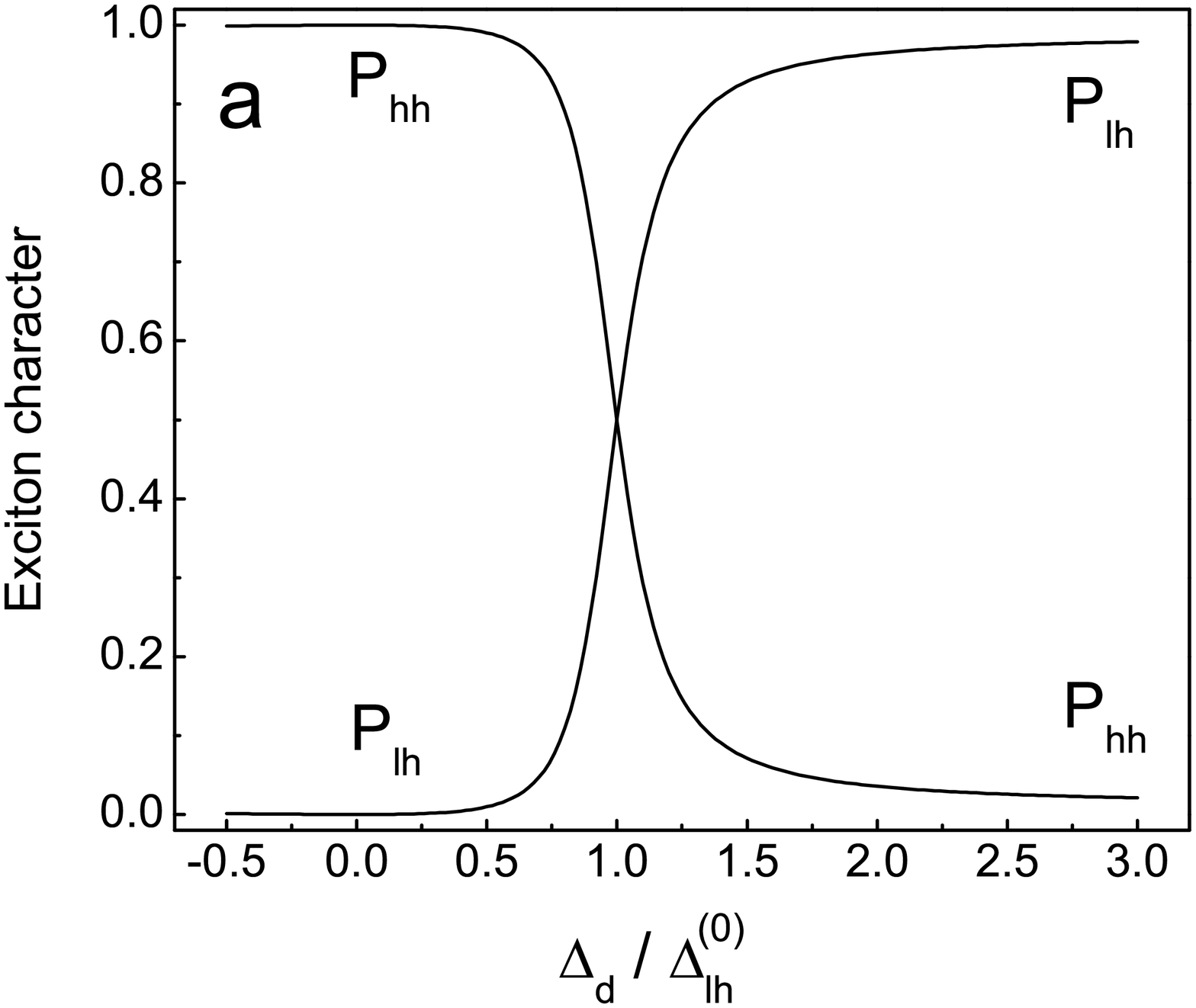}
\includegraphics[scale=.3]{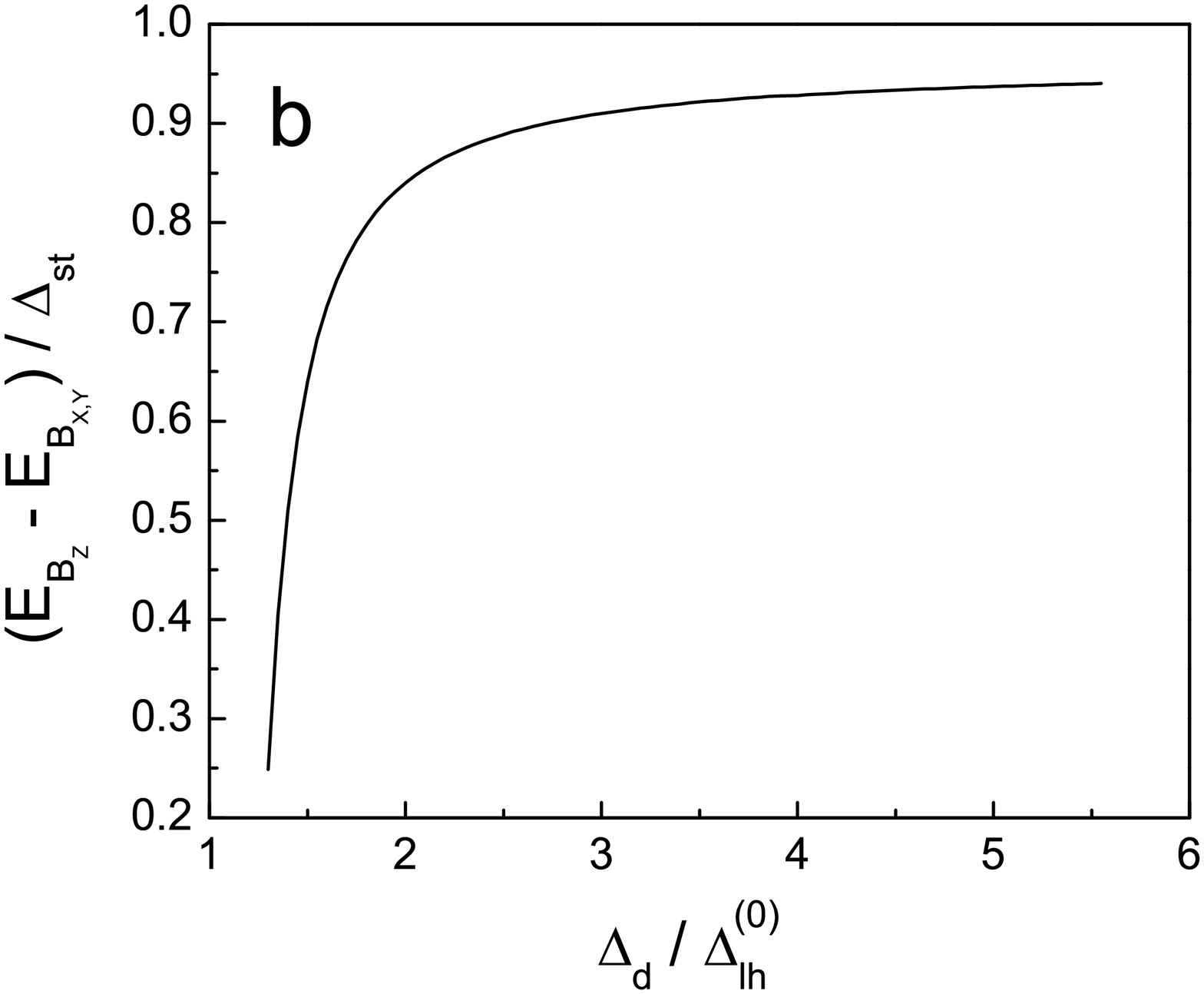}
\caption{Evolution of the exciton characters of the ground state
with in--plane biaxial strain at  $|\gamma| = 0.1$ (a).  Energy
splitting between the optically allowed excitonic states at
$|\gamma| = 0.1$ (b). }{\label{fig1}}
\end{figure}

Simulating the experimental conditions realized in Ref.\cite{Huo},
the excitonic states with $|J_z|=1$ in a strained dot can be
described by a model $(4\times 4)$ Hamiltonian having a
block--diagonal matrix form. Both matrixes, one in the
$\{|-\frac{1}{2};+\frac{3}{2}\rangle\;,
|-\frac{1}{2};-\frac{1}{2}\rangle\}$ basis and another in the
$\{|+\frac{1}{2};-\frac{3}{2}\rangle\;,
|+\frac{1}{2};+\frac{1}{2}\rangle\}$ basis, are identical and
given by
\begin{eqnarray}\label{modHam}
 \left(
\begin{array}{ll}
0  \qquad \qquad \; \gamma\;\Delta_d \\
\gamma^{\star} \; \Delta_d \quad \quad \Delta_{lh}^{(0)} -
\Delta_d
\end{array}
\right)\;.
\end{eqnarray}
Here the energy position of the HH exciton is set to zero and the
strain--induced splitting $\Delta_d$, positive (negative) for the
tension (compression), is determined by the relative deformation
$(\varepsilon_{\parallel} - \varepsilon_{\perp})$. The adjustable
parameter $\gamma$ measures the HH--LH coupling due to the strain
anisotropy, which is expected to be rather weak. The low--energy
states of the bright exciton have still the above structure
Eq.(\ref{Ellipstate}), now with the (absolute value) amplitudes
\begin{eqnarray}\label{amplitudes}
|w_{lh(hh)}| = \frac{1}{\sqrt{2}} \Bigl[1 \mp \frac{1 -
 \delta_d}{\sqrt{(1 - \delta_d)^2 + 4 |\gamma|^2
\delta_d^2}}\Bigr]^{\frac{1}{2}}\;,
\end{eqnarray}
where $\delta_d = \Delta_d/\Delta_{lh}^{(0)}$. In Fig.~\ref{fig1}a
we plot the exciton characters $P_{hh}$ and $P_{lh}$ calculated
from Eq.(\ref{amplitudes}) in a dependence on dimensionless
variable $\delta_d$ at the anisotropic parameter $|\gamma| = 0.1$.
It is seen that while the HH exciton fully dominates in the ground
state for a compressive strain, the exciton character can be
shifted to dominantly LH at a tensile strain. A step--like
switching happens at $\delta_d=1$, where the strain effect
compensates the confinement--induced splitting
$\Delta_{lh}^{(0)}$, and at an increase in tension only twice the
LH exciton character almost completely prevails. This result is
close to that reported for the valence band ground state in a
specific quantum dot calculated in an empirical pseudopotential
based approach in Ref.\cite{Huo}.

Similarly to the bright exciton, the (anisotropic) strain couples
the dark states of the HH exciton with $J_z =|2|$ to the states of
the LH exciton with $J_z=0$. Now, however, the exchange
interaction must be considered explicitly. The corresponding
Hamiltonian represents four--by--four matrix in the
$\{|+\frac{1}{2};+\frac{3}{2}\rangle\;, (2^{-0.5})[-i
|+\frac{1}{2};-\frac{1}{2}\rangle +
|-\frac{1}{2};+\frac{1}{2}\rangle]\;, (2^{-0.5})[+i
|+\frac{1}{2};-\frac{1}{2}\rangle +
|-\frac{1}{2};+\frac{1}{2}\rangle]\;,|-\frac{1}{2};-\frac{3}{2}\rangle\}$
basis. Solving the system makes possible to receive an information
on the fine structure splitting of the exciton ground state. For a
dot with the "mainly LH" exciton ground state ($\delta_d>1$), the
energy splitting between the optically active excitons is
approximated by
\begin{eqnarray}\label{energy}
E_{B_z} - E_{B_{x,y}} \simeq \Delta_{st} \Bigl[1 - \Bigl(\frac{2
\gamma \delta_d}{\delta_d -1}\Bigr)^2 \Bigr]\;.
\end{eqnarray}
Here the energy $E_{B_z}$ and $E_{B_{x,y}}$ refers to the
$z$--polarized exciton and the exciton doublet polarized in the
growth plane of a dot, respectively, and the exchange energy is
$\Delta_{st}$. To illustrate, in Fig.~\ref{fig1}b is shown the
(relative) energy splitting Eq.(\ref{energy}) as a function of
dimensionless parameter $\delta_d$ at $|\gamma|=0.1$ for a dot
with the LH exciton character $P_{lh}\gtrsim 0.88$ (corresponding
to $\delta_d \gtrsim 1.3$). It is seen that the energy distance
between the high--energy exciton $B_z$ and the low--energy doublet
$B_{x,y}$ grows with an increase of a strain and limits to
$\Delta_{st}$ at $\delta_d \gg 1$, as expected for the "pure" LH
exciton \cite{Efros}. These results are similar to experimental
findings and theoretical calculations from Ref.\cite{Huo}.

\begin{figure}
\vspace*{0.5cm}
\begin{center}
\includegraphics[scale=.3]{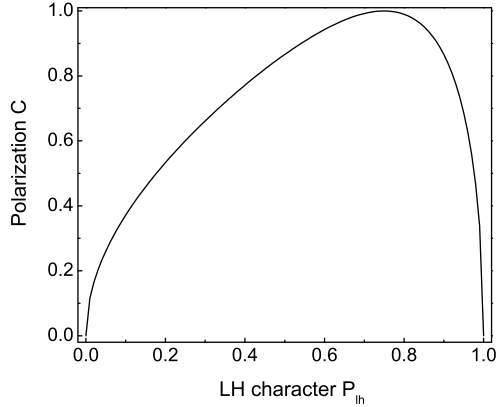}
\caption{Degree of the linear polarization as a function of the LH
exciton character.}{\label{fig2}}
\end{center}
\end{figure}

For the in--plane polarized doublet in a strained dot, in
Fig.~\ref{fig2} we plot a degree of the linear polarization given
by Eq.(\ref{degreeofpol}) as a function of the LH exciton
character $P_{lh}$. Since a contribution of the HH exciton to
recombination is (three times) larger than a contribution of the
LH exciton, the calculated curve is not symmetric with respect to
$P_{lh} \leftrightarrow 1-P_{lh} = P_{hh}$ replacement and the
polarization rate shows a sharp falling at $P_{lh} \rightarrow 1$.
A speed with which the ground state becomes the almost pure LH
exciton depends on a degree of the HH--LH coupling. Indeed,
according to Eq.(\ref{amplitudes}), for the tensile strain the LH
exciton character limits to $P_{lh} \simeq 1 -
\bigl(\varrho/\Delta_{lh}^{(-)}\bigr)^2$ at $\delta_d > 1$, where
$\varrho = |\gamma| E_d$ and $\Delta_{lh}^{(-)} = E_d -
\Delta_{lh}^{(0)} >0$. Obviously, the HH exciton character in this
case is $P_{hh} \simeq \bigl(\varrho/\Delta_{lh}^{(-)}\bigr)^2$.\\
In this regard, an application of results from Ref.\cite{Tonin} to
(at least) the tensile strained dots is, evidently, not correct.
Remember that for the ground state in a quantum dot, the exciton
characters are presented in Ref.\cite{Tonin} by $P_{lh} = \beta^2$
and $P_{hh} = 1 - \beta^2$ with
$\beta=\varrho_s/\sqrt{\Delta_{HL}^2 + \varrho_s^2}$ (formula (1),
where $\Delta_{HL}$ and $\varrho_s$ denotes the valence band
splitting and the coupling amplitude, respectively). These
results, written down in a weak mixing limit, are able to describe
in general the effects of a weak intrinsic anisotropy and/or an
external compressive strain. For tensile strained dots, however,
formal equating of the "mixing parameter" $\beta$ to unity (to
obtain for the polarization rate the desired result $C = 0$) means
a very strong coupling limit and is not adequate to the real
physical scenario described above.

On the contrary, to minimize the optical anisotropy effect in a
strained quantum dot with a dominant light--hole exciton ground
state,  the applied (tensile) strain is required to be highly
isotropic in the growth plane of a dot. Indeed, even if the
exciton ground state has about $95\%$ LH character a degree of the
linear polarization is calculated from Eq.(\ref{degreeofpol}) to
be $C \approx 0.6$. Note that isotropic strain is also desirable
to avoid the strain--induced source of the spin relaxation within
the exciton ground state limiting the
generation of single photons from a dot \cite{Reischle}.\\
Strictly speaking, even in a fully isotropic case the HH and LH
exciton states,  those with the same momentum $J_z = 1 (-1)$, are
coupled by the short--range exchange interaction (evidently, the
heavy-- and light--hole valence band states experience any such
mixing). This kind of a coupling provides a smooth switching from
the HH to LH exciton in the (tensile) strained dot as before, but
preserves the circular polarization of the exciton ground state.
It is possible that such a scenario is relevant for the strained
quantum dot reported in Ref.\cite{Huo}, in which a negligible
degree of the linear polarization $C=0.01$ was measured (inset 3
in Fig. S17). In any case this extremely small polarization rate
points to a very weak LH--HH coupling and the almost pure LH
exciton ground state.


\begin{references}

\bibitem{Huo}
Y.H.Huo et al, Nature Physics {\bf 10}, 46 (2014).


\bibitem{Tonin}
C. Tonin et al, PRB {\bf 85}, 155303 (2012).

\bibitem{Efros}
Al. L. Efros et al, Phys.Rev.B {\bf 54}, 4843 (1996).

\bibitem{Leger}
Y. L$\acute{e}$ger et al, PRB {\bf 76}, 045331 (2007).

\bibitem{Reischle}
M. Reischle  et al,
Phys. Rev. Lett. {\bf 101}, 146402 (2008) and references therein.

\end{references}
\end{document}